# Is Higher Team Gender Diversity Correlated with Better Scientific Impact?

Chengzhi Zhang [*], Jiaqi Zeng, Yi Zhao

Department of Information Management, Nanjing University of Science and Technology, Nanjing, 210094, China

**Abstract**: Collaborative research involving scholars of various genders constitutes a prominent theme in scientific research that has garnered substantial attention. While several studies have investigated the connection between gender-specific collaboration patterns and the scientific impact of paper, the specific gender diversity factors that contribute to enhanced scientific impact remain largely unexplored. In this study, we analyze the correlation between gender diversity and the scientific impact of papers using the examples of Natural Language Processing (NLP) and Library and Information Science (LIS) domains. Our findings reveal three key observations: First, significant gender disparities exist in both NLP and LIS domains, with underrepresentation of female scholars. The gender disparity is more pronounced in the NLP domain compared to the LIS domain. Second, based on papers from the NLP and LIS domains, we find that papers with different gender compositions achieve varying numbers of citations, with mixed-gender collaborations gradually obtaining higher average citation counts compared to same-gender collaborations. Lastly, there is an inverted U-shaped relationship between the gender diversity of paper collaborations and the number of citations received by those papers. Based on the most impactful gender diversity calculations, the ideal gender ratio for NLP and LIS teams within a range where one gender constitutes 5% to 15% of the total number of authors. This paper contributes to the exploration of the most impactful gender diversity in collaborative research and offers insights to guide more effective scientific paper collaboration.

## 1. Introduction

In recent years, a surge in studies exploring gender differences has emerged, fueled by heightened societal awareness of women's experiences in the scientific community. Gender disparity pervades academia, manifesting as social, political, intellectual, cultural, or economic disparities that are rooted in gender or gender identity. This disparity may further encompass variations in access to essential resources, including healthcare, education, economic opportunities, and political freedom, as well as attitudes that can ultimately influence success (Mohammad, 2020a). Gender differences in academia are significant, with data from 2018 showing that globally, only 33.3% of scholars are women, despite a gradual increase in this percentage over time. There are also significant variations in the distribution of female scholars across regions, with Southeast Asia having the lowest percentage of female scholars at 26.3% (Lewis et al., 2021). Huang et al. reported that female authors constituted only 12% of all active authors in 1955, but this proportion has steadily increased over the past century, reaching 35% in 2005, reflecting a growing proportion of female scholars in academia (Huang et al., 2020). This upward trend in female representation may promote greater gender diversity within teams and foster a more inclusive work environment.

[*] Corresponding author, Email: zhangcz@njust.edu.cn

Team science, a subfield within the broader Science of Science (Fortunato et al., 2018), is dedicated to examining and improving the prerequisites, collaborative dynamics, and resultant outcomes of team-based scientific programs (Stokols et al., 2008). Gender diversity plays a pivotal role in harnessing the full potential of each team member, and it has the capacity to expand researchers' perspectives, problem-solving approaches, and fields of inquiry, thereby fostering novel discoveries (Nielsen et al., 2017). While existing research has predominantly explored the relationship between various gender composition patterns and the scientific impact of paper, it is equally important to investigate how the composition of mixed-gender teams influences the scientific impact of their work. For instance, Rogelberg & Rumery (1996) found significant differences in team decision quality across different gender combinations, while Knouse & Dansby (1999) reported that workgroup effectiveness was increased when group gender diversity was at 11-30%. However, a notable research gap exists in comprehending the ideal gender diversity within teams and discerning how this diversity influences the scientific impact of paper. Thus, exploring this topic can improve our understanding of gender diversity in collaboration and promote the effectiveness of team collaboration in academia. Existing theories provide two contrasting views: the information/decision-making perspective posits that group diversity, inclusive of gender diversity, can enhance the scientific impact of paper (Williams & O'Reilly, 1998; Van et al., 2004); on the other hand, social identity theory and self-categorization theory propose that gender diversity may engender conflicts within teams, hampering collaboration efficiency and thereby reducing the scientific impact of paper (HOGGY & Terry, 2000; Trepte & Loy, 2017).

In this paper, we examine the influence of team gender diversity on the scientific impact of papers in two domains, namely Natural Language Processing (NLP) and Library and Information Science (LIS). In this paper, gender diversity refers to the balance of gender within a team. The maximum gender diversity is achieved when males and females are equally represented, with each gender accounting for 50% of the team. The formula for calculating gender diversity is presented in section 3.4. Furthermore, we also pay attention to the gender composition patterns of teams. Based on the gender composition of collaborative teams, there are three distinct composition patterns: all-female collaboration, mixed-gender collaboration, and all-male collaboration. Collaborative research in the NLP domain is widespread. A study was conducted by Mariani et al. analyzing 65,003 articles written by 48,894 different authors in the NLP domain. The research findings revealed that the average number of co-authors per paper increased over time. Additionally, it was observed that 42,471 authors (87% of the total authors) had never published a paper as a single author, indicating a high prevalence of collaboration in NLP research (Mariani et al., 2017). The NLP domain is a male-dominated research area, and the most impactful gender diversity in teams may vary across different domains. In this article, the most impactful gender diversity refers to the gender diversity within a research team that corresponds to the highest citation count achieved by an academic paper. Library and Information Science (LIS), on the other hand, is an interdisciplinary field that has attracted a significant amount of scholarly attention. Early literature indicates that the ratio of male and female scholars in LIS is approximately equal or even slightly higher for women (Herubel, 1992;

Nisonger, 1996; Zhang et al., 2023a). Therefore, we explore the relationship between team gender diversity and paper citation counts based on papers in the LIS domain. Unlike most previous studies, this research aims to identify the range of authorship for each gender that results in higher scientific impact of paper, enriching the existing knowledge on gender differences in team collaboration. Additionally, the proposed methodology can be applied to other domains, contributing to the understanding of the most impactful gender diversity and promoting the advancement of team collaborations.

This paper seeks to answer the following two research questions:

***RQ1***: Do the collaborative patterns of different genders in papers within the NLP and LIS domains correlate with the scientific impact of the papers?

***RQ2***: What is the relationship between gender diversity in team collaboration and higher scientific impact of paper?

The main contributions of this paper are as follows:

We identify the most impactful range of gender ratio in paper collaborations within the NLP domain and LIS domain. Yang et al. (2022) discovered in the medical domain that, compared to same-gender teams, mixed-gender scientific teams tend to produce more innovative papers and achieve higher citation counts. They also identified a positive correlation between maximum gender diversity within a team (with equal representation of males and females) and higher scientific impact of papers (Yang et al., 2022). Nevertheless, there are variations in the authorship patterns across different domains (Larivière et al., 2016), and scholars of different genders may have preferences for specific roles within teams (Larivière et al., 2021). Therefore, it is important to note that the findings from Yang et al. (2022) based on the medical domain may not necessarily be applicable to the NLP and LIS domains. In this study, our aim is to investigate whether similar patterns exist in the domains of Natural Language Processing (NLP) and Library and Information Science (LIS). Utilizing papers from the top three NLP conferences and 14 journals in the LIS domain as our dataset, we examine the impact of fine-grained gender diversity on the scientific impact of papers, measured through two-year citation counts. We explore the effects of varying gender ratios within teams and seek to identify the most impactful gender ratio for collaborative efforts. This finding holds the potential to guide future researchers in selecting their collaborators.

## 2. Related work

A considerable body of research exists concerning gender differences in team collaboration, with an increasing number of studies indicating that such differences can significantly influence team performance (Campbell et al., 2013; Maddi & Gingras, 2021; Yang et al., 2022). In this section, we offer a concise summary of the existing literature, focusing on gender differences in team collaboration and the disparities observed in publication counts and citation rates between genders.

### 2.1 Gender differences in paper collaboration

Collaboration plays a vital role in the advancement of scientific research and positively influences scholars' career development. However, previous studies have indicated that females tend to be less engaged in collaborative projects and publish more single-authored papers compared to males (Boschini & Sjögren, 2007; McDowell & Smith, 1992). For instance, Boschini & Sjögren found a higher proportion of single-authored papers among females compared to males based on an analysis of 4,040 articles published in economics journals between 1991 and 2002 (Boschini & Sjögren, 2007). Additionally, Jadidi et al. (2018) reported that females are less likely to participate in long-term research collaborations. Nonetheless, some studies present contrasting findings (Kretschmer et al., 2012; Ozel et al., 2014; Fell & König, 2016). For example, Fell & König discovered that female researchers exhibit higher involvement in team collaborations than their male counterparts in the field of psychology, with a paper collaboration rate of 86% for females and 80% for males (Fell & König, 2016). Pinheiro et al. found that the presence of female authors enhances the interdisciplinarity of papers, regardless of whether self-citations are taken into account in the measurement of interdisciplinarity (Pinheiro et al., 2022).

Numerous scholars have conducted research on gender homogeneity in team collaboration, which refers to the tendency of individuals to interact with others of the same gender and is prevalent in both natural and human social relationships (Fu et al., 2012). Several studies have demonstrated that different patterns of gender composition can impact the citation count of papers (Campbell et al., 2013; Maddi & Gingras, 2021). For example, Campbell et al. analyzed the citation numbers of papers authored by individuals of the same gender compared to those authored by individuals of different genders, using collaboration teams from the National Center for Ecological Analysis and Synthesis (NCEAS). They found that gender-heterogeneous authorship teams received, on average, 34% more citations per paper than gender-uniform authorship teams (Campbell et al., 2013). From an information and decision-making perspective, gender diversity within a collaborative team can introduce diverse knowledge, skills, and perspectives, thereby enhancing the team's expertise and problem-solving abilities, leading to positive outcomes (Williams & O'Reilly, 1998; Van et al., 2004). However, theories such as social identity theory and self-categorization theory, rooted in social psychology, suggest that individuals tend to categorize themselves into specific groups based on factors such as gender or nationality before engaging in social comparisons. Consequently, team diversity may result in increased conflict, reduced communication, and decreased team cohesion, all of which can negatively impact team performance (HOGGY & Terry, 2000; Trepte & Loy, 2017).

A substantial body of research has investigated the influence of gender diversity on team performance. Kanter's influential framework categorizes gender-specific groups based on the proportion of males and females, identifying four types of groups: uniform (0/100), skewed (5/95-15/85 and 20/80-35/65), and balanced (40/60-50/50). These different levels of gender diversity can have distinct effects on team performance (Kanter, 1977; Kanter, 1987). Studies have shown that work groups with a homogeneous gender ratio (0/100) may suffer from a lack of innovation and subpar performance. Conversely, teams with higher gender diversity levels (40/60-50/50) tend to be divided into roughly equal groups of males and females. However, when a minority group (e.g., females) is overrepresented, it can be perceived as a threat by the majority group (e.g., males), leading to increased conflict and diminished teamwork performance (Allport & Atkinson, 1954; Blalock, 1967). From both an information/decision-making perspective and social categorization theory, the level of team gender diversity can dynamically impact organizational performance, with a certain level of gender diversity being most beneficial for collaborative teams. Numerous studies have investigated the impact of different gender composition patterns on scientific impact of paper in academic paper co-authorship (Campbell et al., 2013; Maddi & Gingras, 2021; Yang et al., 2022). For instance, Maddi and Gingras found that mixed-gender collaborative publications received more citations than publications from same-gender collaborative teams in economics and management journal papers (Maddi & Gingras, 2021). Similarly, Yang et al. (2022) found that publications with mixed-gender collaborations tended to be more novel and influential than publications from same-gender groups of the same size. Some scholars have also examined the effect of mixed-gender collaboration on scholars' research performance. For example, Zhang et al. (2024) found that moderate cross-gender collaboration contributes to disruptive innovation. There are no significant physiological differences between men and women in terms of innovation; the key is to find gender balance in collaboration. Shen et al. (2022) found that collaborating with more cross-gender scholars had a significant negative impact on the research performance of female scholars and a positive impact on male scholars. This negative effect on female scholars was further amplified when partners were at higher academic levels. However, fewer studies focus specifically on the effect of gender diversity within teams on team performance. Existing studies related to gender differences in paper co-authorship are summarized in Table 1.

In summary, studies have highlighted disparities in the involvement of different genders in team collaboration. Research suggests that mixed-gender collaborations often yield a higher scientific impact than same-gender collaborations. However, the most impactful level of gender diversity in team collaboration remains unclear. This study aims to investigate this relationship based on papers from the domains of NLP and LIS.

**Table 1. Literature related to gender differences in paper collaboration**

| Year | Authors | Domain | Size of Dataset | Main Findings |
|---|---|---|---|---|
| **2007** | Boschini & Sjögren (2007) | Economics | 4,040 articles published in three top economics journals between 1991 and 2002 | Females are generally less involved in collaborative projects than males, and their publications are more likely to be single-authored papers |
| **2013** | Campbell et al. (2013) | Ecology | NCEAS Citation Data for Papers Authored by Mixed-Gender and Same-Gender Research | Gender-heterogeneous authorship teams receive an average of 34% more citations per paper than gender-uniform authorship teams |
| **2016** | Fell & König (2016) | Psychology | Publications written by 4,234 (45% female) industrial organizational psychologists from 1948 to 2013 | The paper collaboration rate was 86 % for females and 80 % for males, female researchers were more involved in team collaborations than men, and same-gender collaboration is more common than cross-gender collaboration |
| **2017** | Jadidi et al. (2018) | Computer Science | Collaboration models for over one million computer scientists | Female scholars less willing to engage in long-term research collaborations |
| **2021** | Maddi & Gingras (2021) | Economics | 153,667 articles in economics journals and 163,567 articles in management journals published between 2008 and 2018 on the Web of Science | Mixed-gender co-publications receive more citations than same-gender co-publications |
| **2022** | Yang et al. (2022) | Medicine | More than 6.6 million research publications in more than 15,000 medical science journals from 2000 to 2019 | Mixed-gender collaborative publications are more innovative and impactful, there is a positive correlation between the maximum gender diversity within the team and the scientific impact of papers. |
| **2022** | Hajibabaei et al. (2022) | Artificial Intelligence | 39,679 articles in the field of artificial intelligence | The proportion of mixed-gender collaborations in artificial intelligence is generally on the rise |
| **2022** | Shen et al. (2022) | Computer Science | 2,956,499 journal and conference papers published by 1,338,957 unique scholars | Mixed-gender collaboration has a significant negative impact on the research performance of female scholars and a positive impact on male scholars. When partners are at a higher academic level, the negative impact on female academics is reinforced |
| **2024** | Zhang et al. (2024) | Physics | 136,791 publications from the American Physical Society (APS) | Compared to single-gender teams, moderate gender diversity in collaboration has greater potential to generate disruptive knowledge. There is no significant difference between men and women in terms of innovation; the key is to find gender balance in collaboration. |

### 2.2 Gender disparities in the number of publications and citation counts

Gender differences in the number of publications refer to the discrepancies in the number of papers published by scholars of different genders. Numerous studies have investigated this topic. For example, Paul-Hus et al. (2015) analyzed 1,059,939 papers from Russia indexed by the Web of Science from 1973-2012 and found that female research output accounted for between 20% and 30% of all output. Frietsch et al. (2009) analyzed the gender and paper output of 490,244 scholars from 14 countries/regions via the Scopus database. They observed that although women's contribution to research output has gradually increased over time in most countries/regions, it remains relatively low at 20%-25%. Lerman et al. (2022) examined gender differences among prominent scholars elected to the National Academy of Sciences and found that female scholars were in the minority. Specifically, psychology had the highest percentage of women (42%), followed by sociology (22%), astronomy (18%), chemistry (13%), economics (12%), computer science (11%), and physics (8%).

Gender differences in the paper citation index refer to the variation in the number of citations of papers published by scholars of different genders. Several studies have documented differences in the citation counts of papers authored by scholars of different genders (Larivière et al., 2013; Chatterjee and Werner, 2021; Teich et al., 2022; Liu et al., 2022). For example, Larivière et al. (2013) analyzed a dataset of over 5 million research papers and found that in countries with high paper output, articles dominated by women received fewer citations than those dominated by men (Larivière et al., 2013). Teich et al. (2022) focused on the field of physics and discovered that papers written by women were cited less frequently than papers authored by men. In high-impact medical journals, articles authored by females as first or last authors were also found to receive fewer citations than those authored by male first or last authors, highlighting the potential impact of gender differences on women's career success and equity in academia (Chatterjee & Werner, 2021).

Furthermore, the COVID-19 pandemic has exacerbated gender inequalities in science, with a decline in the proportion of female scholars and a decrease in citation counts for papers authored by women (Liu et al., 2022).

Overall, significant disparities exist in the publication output between scholars of different genders, with men tending to produce more research papers than women across most fields of study. Furthermore, numerous studies have identified a gender gap in paper citations, with papers authored by female scholars generally receiving fewer citations compared to their male counterparts. These findings highlight the impact of gender on academic performance and suggest that the gender composition of authorship can influence research outcomes.

Numerous scholars have examined gender differences in paper authorship, with findings typically indicating that men tend to publish more papers and receive more citations compared to women. Based on research in the medical domain, Yang et al. (2022) found that papers produced by mixed-gender scientific teams are more innovative and garner higher citation counts compared to papers from same-gender teams. They also observed that as a team's gender balance approaches a 50/50 split between women and men, the positive association between gender-diverse teams and innovation and impact intensifies. Our study differs from theirs in several ways. First, our focus is on the Natural Language Processing (NLP) and Library and Information Science (LIS) domain, which is different from the medical domain, gender differences can vary across different domains (Thelwall, 2020; Chan & Torgler, 2020). Additionally, we aim to identify the gender ratio range for NLP and LIS teams associated with high scientific impact. Our findings differ from the study conducted by Yang et al. (2022). Our results suggest an inverted U-shaped relationship between gender diversity and scientific impact, rather than a direct correlation between maximum gender balance (equal proportions of men and women) and maximum scientific impact. This contributes new insights into the relationship between team gender diversity and the scientific impact of papers.

## 3. Methodology

### 3.1 Framework of the Study

In this study, we examine gender diversity among authors in the NLP (Natural Language Processing) and LIS (Library and Information Science) domain. We analyze various gender composition patterns in team collaboration and investigate the impact of gender diversity on scientific impact. The scientific impact of paper is measured using the two-year citation count of the papers. Our goal is to explore a potential correlation between gender diversity in team collaboration and the scientific impact of paper through correlation analysis. This study aims to enhance researchers' understanding of gender differences in academia and team collaboration, providing valuable insights and guidance for researchers forming teams and conducting team science-related research.

The framework of this paper is presented in Figure 1. In the data pre-processing stage, we collect papers in the NLP and LIS domains, record the authors' information for each paper, and obtain the number of citations for each paper. To identify the authors' gender, we use name-gender lists for author gender-matching identification. This is followed by the use of gender recognition tools and manual search identification to ascertain the gender of authors in our dataset. In the analysis stage, our primary focus is on exploring the trends in the proportion of collaborative papers over time. Then, we analyze the results in two aspects: 1) gender composition patterns in paper authorship (all-male, all-female, mixed-gender), 2) fine-grained gender diversity in collaboration and its correlation with the scientific impact of paper.

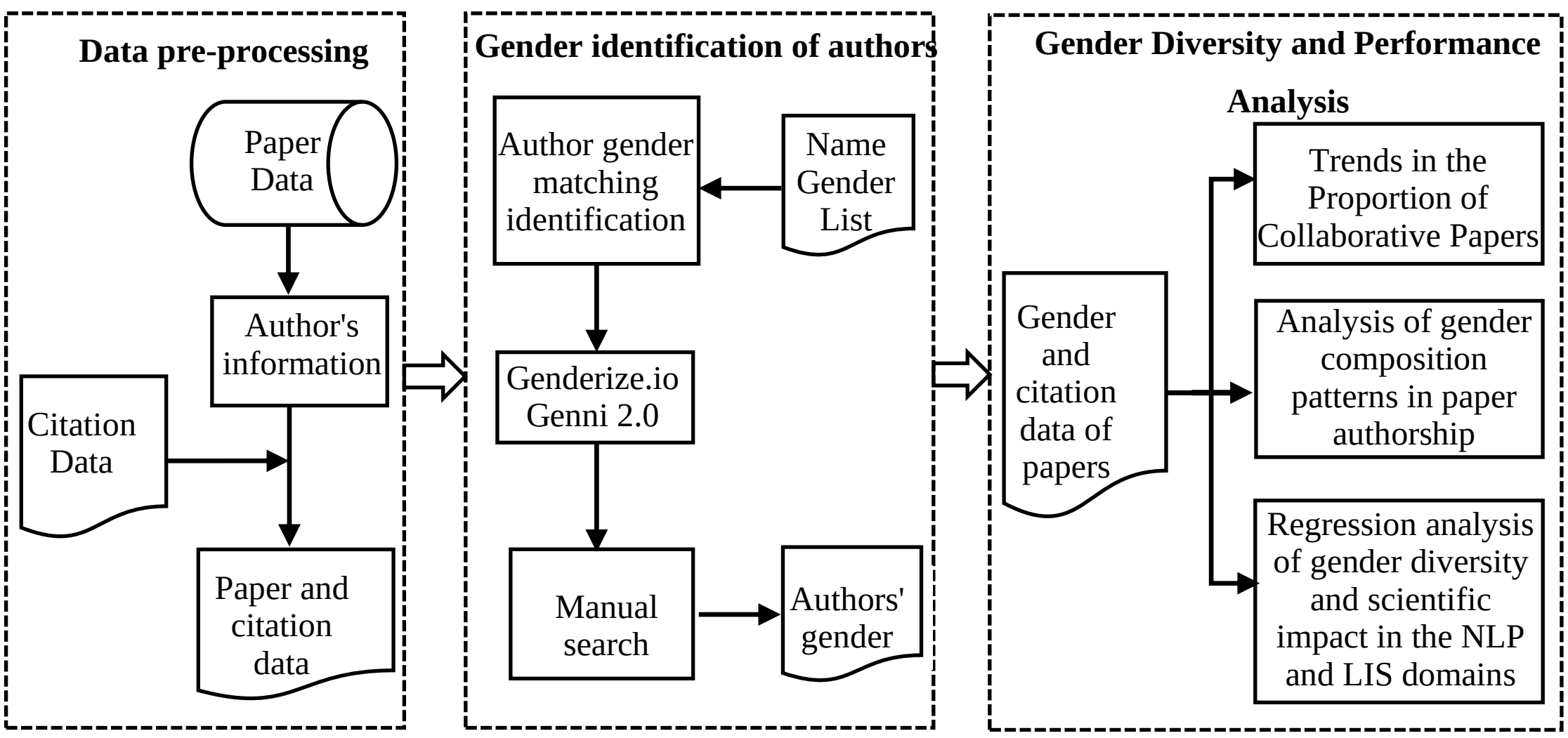

**Figure 1. Framework of this study**

### 3.2 Data collection and processing of authors and citations

(1) Author data of papers

This paper presents an analysis of the NLP (Natural Language Processing) and LIS (Library and Information Science) domains. For the NLP domain, this study focuses on papers from the top three conferences in the NLP domain: *Annual Meeting of the Association for Computational Linguistics* (*ACL*), *North American Chapter of the Association for Computational Linguistics* (*NAACL*), and *Conference on Empirical Methods in Natural Language Processing* (*EMNLP*). The NLP domain is a broad interdisciplinary field encompassing work in language and computation, drawing influence from artificial intelligence, computer science, linguistics, psychology, and social sciences. It comprises approximately 30% female scholars, which aligns with the global percentage of female scholars (Mohammad, 2020a). In the computer science domain, authors often showcase their highly innovative and outstanding research through conference publications rather than journals (Wang & Zhang, 2020). *ACL*, *EMNLP*, and *NAACL* represent the most prestigious conferences in the NLP domain, dedicated to addressing computational challenges related to human language (Zhang et al.,2023b). Hence, this study utilizes data from these premier conferences. We obtain the authors, titles, and abstracts of all main conference papers from *ACL*, *NAACL*, and *EMNLP* spanning from 1979 to 2021 through the ACL Anthology website[1]. Additionally, we collect information regarding the authors' order and affiliations for each paper. NLP dataset comprises a total of 15,337 papers, including 7,205 *ACL* papers with 23,345 authors, 2,696 *NAACL* papers with 9,115 authors, and 5,436 *EMNLP* papers with 20,081 authors. Altogether, NLP dataset encompasses 52,541 authors.

For the LIS (Library and Information Science) domain, this paper refers to the journal directory selected by Jarvelin and Vakkari (2022) research, which includes a total of 31 journals. In addition, 14 internationally recognized and authoritative LIS core journals are chosen as the data source for

[1] https://aclanthology.org/

LIS domain research, based on the JCR-LIS journal ranking in top three quartiles(Q1~Q3). The 14 journals include *Aslib Journal of Information Management* (*AJIM*), *College and Research Libraries* (*CRL*), *Information Processing and Management* (*IPM*), *Information Technology and Libraries* (*ITL*) , *International Journal of Information Management* (*IJIM*), *Journal of Documentation* (*JOD*), *Journal of Information Science* (*JIS*), *Journal of Librarianship and Information Science* (*JLIS*), *Journal of the Association for Information Science and Technology* (*JASIST*), *Library and Information Science Research* (*LISR*), *Library Quarterly* (*LQ*), *Online Information Review* (*OIR*), and *Scientometrics* (*SCIM*). This paper collects research papers from 14 journals spanning the years 1990 to 2021. It excludes editorial articles, book reviews, and other non-research papers. In total, there are a total of 23,957 research papers obtained in the LIS domain. The distribution of papers and authors is presented in Table 2.

**Table 2. Distribution of papers and authors**

| Domain | Conferences or Journals | Total Authors | Total Papers |
|---|---|---|---|
| **NLP** | *ACL, NAACL, EMNLP* | 52,541 | 15,337 |
| **LIS** | *SCIM, JASIST, IPM, IJIM, TEL, OIR, JIS, JOD, AJIM, CRL, JLIS, LISR, ITL, LQ* | 57,052 | 23,957 |

In the NLP and LIS domains, it is customary for the first author to be a junior scientist responsible for writing the paper, while the last author is typically a senior researcher who conceptualizes or provides funding for the study. NLP dataset comprises a total of 52,541 authors, with 11,672 (22.2%) being female and 40,869 (77.8%) being male. Table 3 presents the gender ratios for all authors, first authors, and last authors. In general, the first author of a paper is considered the most significant contributor. Among the 15,337 papers included in NLP dataset, 3,564 (23.2%) have female first authors and 11,773 (76.8%) have male first authors. For the last authors, 3,335 (21.7%) papers have female last authors and 12,002 (78.3%) have male last authors. The LIS dataset consists of a total of 50,193 authors, of which 17,838 (35.5%) are female and 32,355 (64.5%) are male. Among the first authors, 8,197 (37.4%) are female first authors, while 13,703 (62.6%) are male first authors. Regarding the last authors, 7,564 (34.5%) papers have a female last author, while 14,336 (65.5%) papers have a male last author.

The analysis of our dataset reveals that there is underrepresentation of female authors in both NLP and LIS domains, with a larger gender disparity observed in the NLP domain compared to the LIS domain. Table 3 provides comprehensive information on the gender ratios of authors.

**Table 3. Gender ratio of authors**

| Domain | Gender | Number of authors （percentage） | Number of first authors （percentage） | Number of last authors （percentage） |
|---|---|---|---|---|
| NLP | Female | 11,672（22.2%） | 3,564（23.2%） | 3,335（21.7%） |
| | Male | 40,869（77.8%） | 11,773（76.8%） | 12,002（78.3%） |
| | Total | 52,541（100%） | 15,337（100%） | 15,337（100%） |
| LIS | Female | 17,838（35.5%） | 8,197（37.4%） | 7,564（34.5%） |
| | Male | 32,355（64.5%） | 13,703（62.6%） | 14,336（65.5%） |
| | Total | 50,193（100%） | 21,900（100%） | 21,900（100%） |

(2) Citation data of papers

This paper measures the scientific impact of papers through citation data. However, it is worth noting that citation databases like *Web of Science* do not provide complete coverage for all the papers included in this study. Many researchers have turned to *Google Scholar to* retrieve citation information (Mohammad, 2020b) (Martín-Martín et al., 2018), as there exists a strong correlation between citation data from *Web of Science* and *Google Scholar* (Martín-Martín et al., 2018). Additionally, considering the emphasis of this study, it focuses on NLP conference papers and LIS journal papers, *Google Scholar* offers highly comprehensive citation data (Martín-Martín et al., 2018). It encompasses citations from various sources, including journals, conferences, and more, providing a better reflection of the scientific impact of the papers. Consequently, we retrieve the citation data for all the papers within our dataset by searching their titles and author names on the *Google Scholar* website. To account for the cumulative effect of citation counts over time, we only consider the number of citations received within two years following each paper's publication as a measure of scientific impact. Specifically, we begin by retrieving the article IDs from the *Google Scholar* website using the authors' names and paper titles. Subsequently, we obtain the citation data for the papers within two years by using a combination of article IDs and the specified citation time frame in the query.

### 3.3 Author gender identification

Since the *ACL*, *NAACL*, *EMNLP* conferences, and the 14 LIS journals do not provide gender information of authors, it is necessary to determine the gender of the authors. Common methods of gender identification include manual annotation, automatic matching with existing name-gender lists, and determining gender based on first names, countries, and other relevant information.

It is important to note that different regions have distinct naming conventions, which can impact the correlation between names and gender. For example, the correlation between names and gender in Asia is generally lower compared to that in Europe and America. To ensure accurate gender identification, we employ a combination of name-gender list matching identification, the Genderize.io and Genni 2.0 tool, and manual search identification (Zhang et al., 2023a). For specific gender recognition methods and the assessment of gender recognition accuracy, please refer to the Supplemental Material. The gender of all authors in NLP papers has been determined. For LIS papers, due to the presence of abbreviations in the names of many authors, making it impossible to identify their gender, we excluded 2,057 papers with incomplete gender recognition. The remaining LIS dataset consists of 21,900 papers.

### 3.4 Variable and regression models

(1) Variable

The measure of a team's scientific impact is based on their paper's two-year citation count. Given the potential for a paper to acquire zero citations within a span of two years, we augment the citation count by one and calculate its natural logarithm, employing ln(citation+1) as the dependent variable for our regression analysis.

The Blau-index serves as our gauge of gender diversity within the team. We employ the Blau-index, which is a heterogeneity index proposed by Blau (1977) to measure gender diversity based on gender ratios. The Blau-index is calculated as (Formula 1), where Pi denotes the proportion of each gender in relation to the total number. The Blau-index is based on a ratio or continuous scale, and the index increases as men and women become more equally represented in the team (Blau, 1977).

$$Blau - index = 1 - \Sigma Pi^2 \quad (1)$$

When assessing gender diversity, the Blau-index ranges from 0 (0/100 gender ratio), indicating a state of gender homogeneity, to 0.5 (50/50 gender ratio), representing the maximum level of gender diversity.

We observe that factors such as team size, number of affiliated institutions, institutional ranking, international collaboration, publication year, and topic of the paper are all associated with the citation count of the paper. The explanations for these variables are presented in Table 4. Table S1 and Table S2 displays the Pearson correlation coefficients for these variables. The low correlation coefficients between the independent variables indicate that multicollinearity among them can be disregarded. Hence, in the subsequent regression analysis, we include these control variables to account for their potential effects on the scientific impact of the paper.

**Table 4. Variable explaination**

| Type | Variable | Calculation Method | Symbol |
|---|---|---|---|
| Paper Characteristics | Team Gender Diversity | The Blau-index of gender diversity within the team, the specific calculation methods can be found in Section 3.4. | B |
| | Team Size | Number of team members in the paper. For teams with fewer than 6 members, set "size" to 0 to indicate small teams, while teams with 6 or more members are categorized as large teams with T=1 (Yang et al., 2022). | Size |
| | Institutional Ranking | Set to 1 when a paper has institutions ranked among the top 500 institutions. NLP domain papers : For university ranking, we utilize the Computer Science Rankings, while for company ranking, we refer to the rankings from a website that collects company market values (CompaniesMarketCap.com ). For LIS domain papers, the rankings are based on the top 500 rankings from the Times Higher Education.[2] | Ranking |
| | Number of Institutions | The number of institutions the authors of the papers are affiliated with. | Institution |
| | International Collaboration (Yes/No) | We use the country affiliation of the authors' institutions to determine whether it is an international collaboration. When a paper's authors are engaged in multinational collaboration, the value of the "international" variable is set to 1. | International |
| | Topic of the Paper | Topics of papers obtained using the Top2Vec model. Top2Verc topic model categorizes NLP papers into 10 | Topic |

[2] https://www.timeshighereducation.com/world-university-rankings/2024/world-ranking

| | | topics based on their titles and abstracts, while LIS papers are classified into 22 topics. (Angelov, 2020) | |
|---|---|---|---|
| | Publication Year | Publication Year of the Paper | Year |
| **Scientific Impact** | Paper Citations | Add 1 to the number of citations received by the paper over a two-year period, and then take the logarithm. | ln (citation+1) |

To investigate the team size of papers in the NLP and LIS domains, we analyze the number of papers with different team sizes, as shown in Figure 2. Figure 2(a) represents the number of papers by team size in the NLP domain. The figure illustrates that the majority of collaborative papers in the NLP domain consist of teams with 2, 3, or 4 authors, while there are relatively fewer teams with more than six authors. In the LIS domain, Figure 2(b) demonstrates a relatively higher prevalence of single-authored papers, while most collaborative papers typically involve 2, 3, or 4 authors. Teams consisting of six or more authors are less common in this domain. By examining the mean average B values across teams of varying sizes, we found that for papers in the domains of LIS and NLP, gender diversity generally exhibits an increasing trend as team size grows.

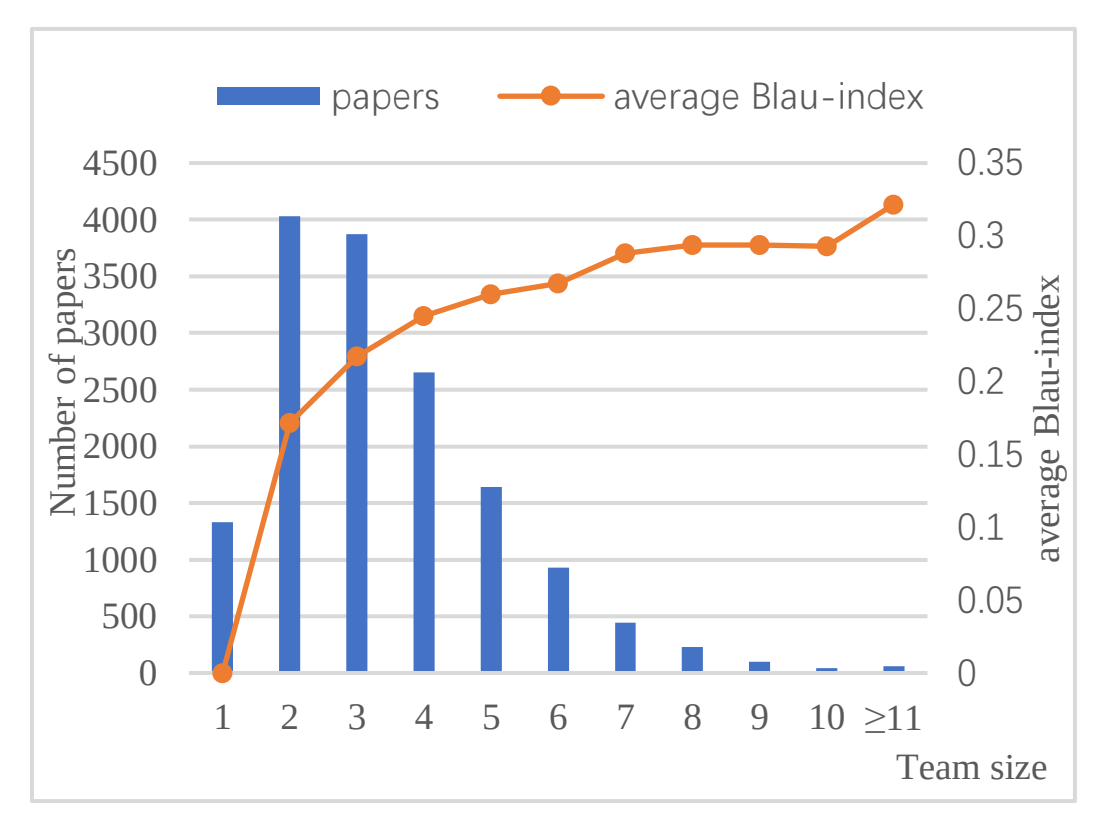

**(a) Number of papers and average Blau-index by team size in the NLP domain**

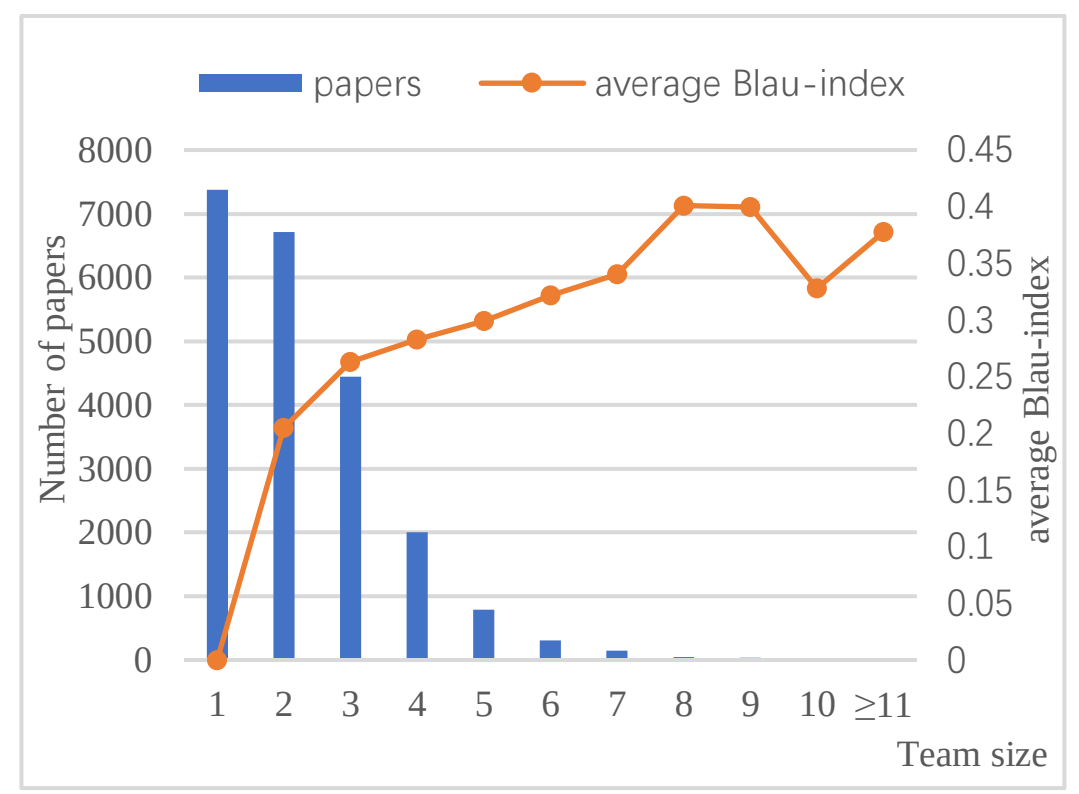

**(b) Number of Papers and average Blau-index by Team Size in the LIS domain**

**Figure 2. Number of Papers and average Blau-index by team size**

Since this study considers the relationship between gender diversity in collaborative papers and citation counts, our primary focus is on collaborative papers. After removing 1,333 single-authored papers from the NLP domain, there remain 14,004 collaborative papers. In the LIS domain, after excluding 7,372 single-authored papers, there are 14,528 collaborative papers remaining. Descriptive statistics for gender diversity (B values) in the NLP and LIS domains can be found in the Table 5.

**Table 5. Descriptive statistics for gender diversity (Blau-index) in the NLP and LIS domains**

| Domain | Variable | Obs | Mean | Std. Dev. | Median | Min | Max |
|---|---|---|---|---|---|---|---|
| NLP | Blau-index | 14004 | .223 | .218 | .278 | 0 | .5 |
| LIS | Blau-index | 14,528 | .244 | .230 | .375 | 0 | .5 |

(3) Regression models

We conduct multiple linear regressions to explore the relationship between gender diversity and the scientific impact of paper. Considering the potential for both positive and negative effects of team gender diversity on scientific impact of paper, we hypothesized that team gender diversity might exhibit a quadratic relationship with scientific impact of paper. Therefore, we included a quadratic term ($B^2$) for the team gender diversity variable (B) in our regression analysis. Our regression model is as shown in Equation 2. Table 4 presents the variables utilized in the regression analysis. $Year_{FE}$ represents the fixed effect of the publication year of the paper, while $Topic_{FE}$ represents the fixed effect of the topic of the paper.

$$\ln(\text{citation}+1) = \alpha_0 + \alpha_1 B^2 + \alpha_2 B + Size + Ranking + Institution + International + Topic_{FE} + Year_{FE} + \varepsilon \quad (2)$$

## 4. Result Analysis

The evolution of science is inextricably linked to collaboration. Thus, an in-depth examination of the trends in paper collaboration within scientific research endeavors, along with the gender disparities, gender composition patterns, gender diversity, and their subsequent effects, holds paramount significance in comprehending the dynamics of scientific research collaboration. In this endeavor, we examine the proportion and trends of collaborative papers and explore the differences in gender composition patterns of paper collaboration. Moreover, we investigate whether there is a correlation between gender diversity in paper collaboration and the scientific impact of the papers. We aim to identify the ideal gender diversity that may contribute to higher scientific impact in team collaborations.

### 4.1 Trends in the proportion of collaborative papers over time

We define collaborative papers as those with two or more authors, which account for 91.3% of the NLP dataset and 66.3% of the LIS dataset. To analyze the trend of collaborative papers over time, we calculated the ratio of single-authored papers to collaborative papers for different years, as shown in Figure 3. Figure 3(a) represents the percentage of single-authored and multi-authored papers in the NLP domain over the years, while Figure 3(b) illustrates the percentage of single-authored and multi-authored papers in the LIS domain over the years.

Overall, Figure 3 indicates that for the NLP domain, the proportion of single-authored papers exceeded that of collaborative papers from 1979 to 1986. However, starting from 1986, the proportion of collaborative papers has been increasing. Nevertheless, even during this period, the proportion of single-authored papers remained higher than that of collaborative papers. Since 1996, the NLP domain has witnessed a continuous decline in the proportion of single-authored papers, while the proportion of collaborative papers has steadily increased. In recent years, collaborative papers account for 99% of all NLP papers. In contrast, for the LIS domain, the proportion of collaborative papers has been consistently increasing since 1989, with approximately 80% of papers being collaborative in recent years. When comparing both domains, the collaboration rate of papers at the top three NLP conferences is higher than that of journal papers in the LIS domain. As paper collaboration becomes an integral aspect of the contemporary academic landscape, it is crucial to recognize this trend and explore more effective and scientific patterns of collaboration to foster progress within the academic community.

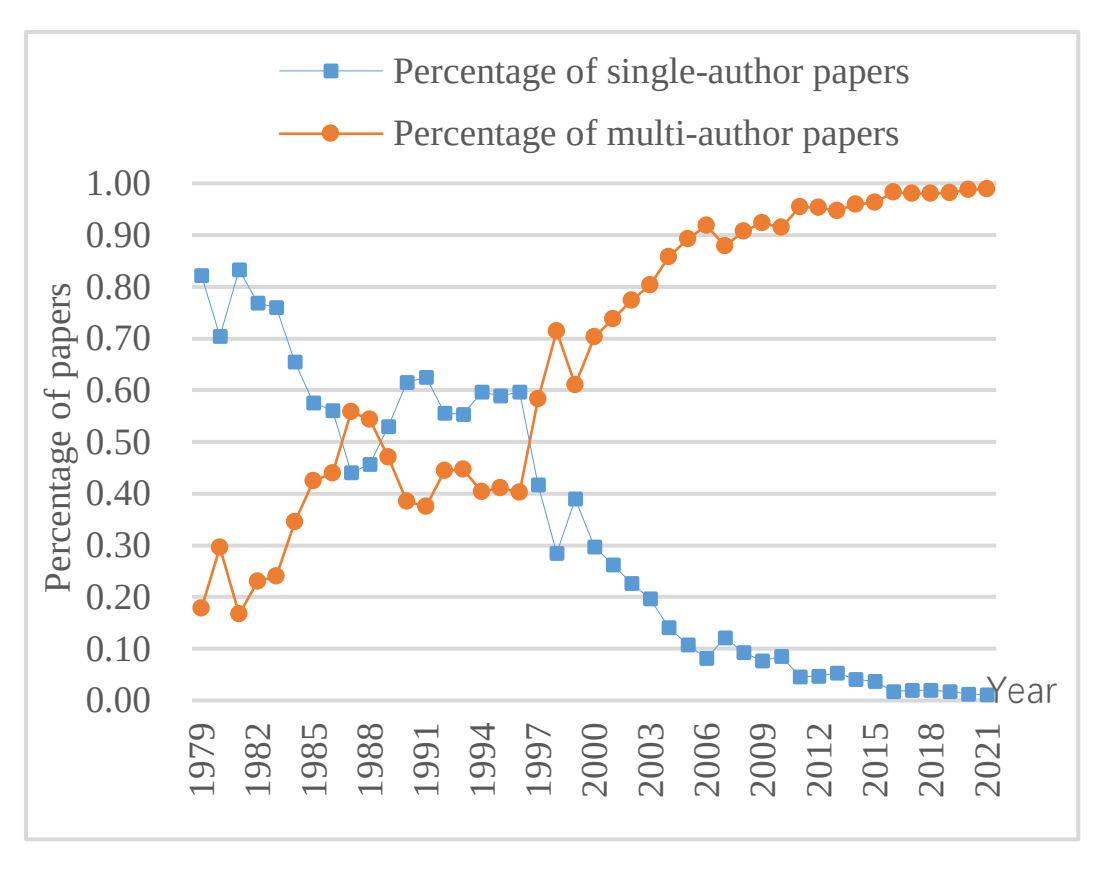

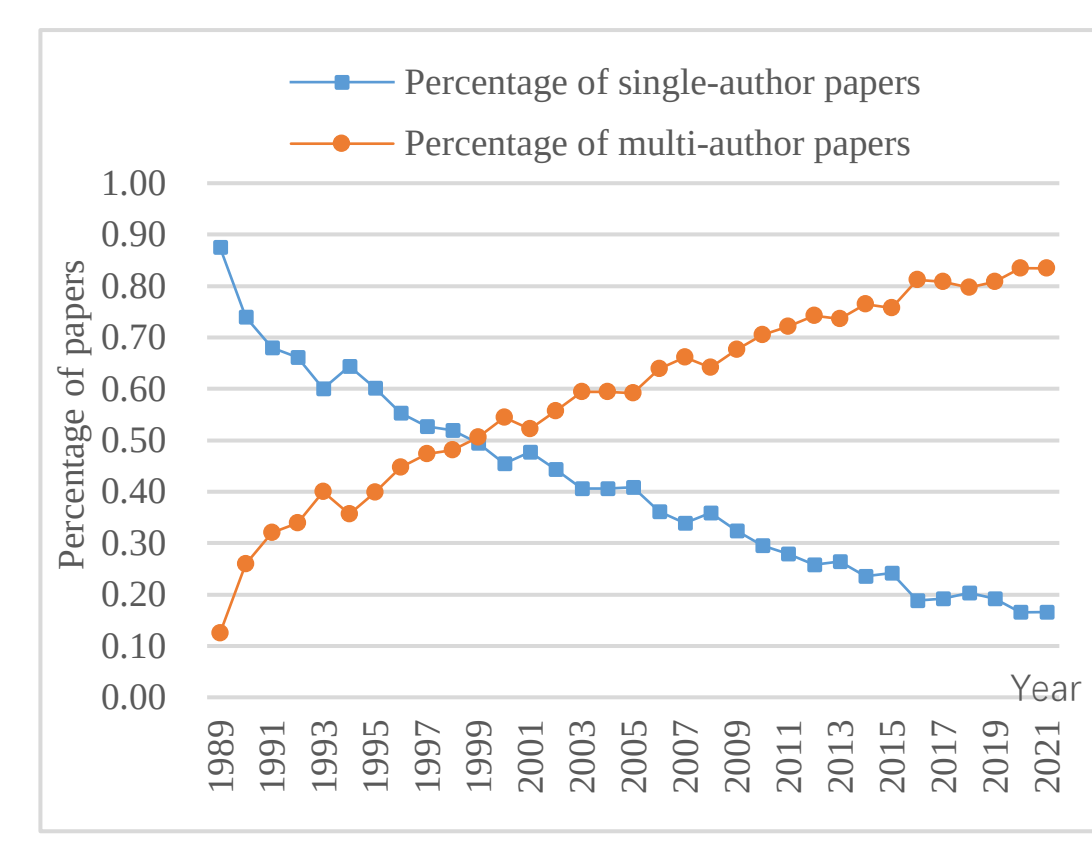

**(a) Percentage of single-authored and multi-authored papers in the NLP domain**

**(b) Percentage of single-authored and multi-authored papers in the LIS domain**

**Figure 3. Percentage of single-authored and multi-authored papers by year**

## 4.2 Different gender composition patterns in papers collaboration

Based on the analysis above, it is clear that collaboration plays a significant role in academic papers. In order to delve deeper into gender differences in composition patterns, we categorize paper collaborations into three groups: all-female collaboration, mixed-gender collaboration, and all-male collaboration. We examine the proportion trends of these three gender composition patterns.

After excluding single-authored papers, the total number of collaborative papers in our NLP dataset is 14,004, while in the LIS dataset, it is 14,528. To examine the gender composition patterns in multi-author papers, we analyze the number and proportion of different collaboration patterns in both databases. The results in Table 6 indicate that in the NLP database, there are 373 all-female papers, accounting for 2.7%, 6,266 all-male papers, accounting for 44.7%, and 7,365 mixed-gender papers, accounting for 52.6% of the total number of multi-author papers. For the LIS database, there are 1,686 all-female papers, accounting for 11.6%, 5,034 all-male papers, accounting for 34.7% of the total number of multi-author papers, and 7,808 mixed-gender papers, accounting for 53.7%. These findings suggest that mixed-gender collaborative papers constitute the largest proportion among multi-author papers, while the number of all-female collaborative papers is relatively low.

**Table 6. Proportion of different gender composition patterns**

| Domain | Gender composition patterns | # papers(percentage) |
|---|---|---|
| NLP | All-Female | 373（2.7%） |
| | All-Male | 6,266（44.7%） |
| | Mixed-gender | 7,365（52.6%） |
| LIS | All-Female | 1,686（11.6%） |
| | All-Male | 5,034（34.7%） |
| | Mixed-gender | 7,808（53.7%） |

Figure 4 displays the proportion of papers with different gender patterns in the NLP (a) and LIS (b) databases over the years. For the NLP data, prior to 2000, the limited number of papers resulted in significant variations in the proportions of the three collaboration patterns. Notably, the proportion of all-female collaborative papers (FF) has remained consistently low without any apparent upward trend over the years. Before 2015, the proportion of papers with all-male collaboration (MM) generally exceeded that of the other patterns, but it has been declining since then. Conversely, the proportion of mixed-gender collaborative papers (FM) has exhibited a steady increase since 2015, becoming the most common pattern in recent years. For the LIS data, the proportion of all-female collaborative papers (FF) has consistently been the lowest over the years. Since 1991, the proportion of mixed-gender collaboration papers (FM) has been the highest and has shown an upward trend. The increasing trend of mixed-gender collaborative papers in both the NLP and LIS databases indicates that interdisciplinary collaboration is becoming more prevalent in scientific research. Additionally, it indicates that female authors are expanding their scientific influence by participating in male-dominated collaborative groups.

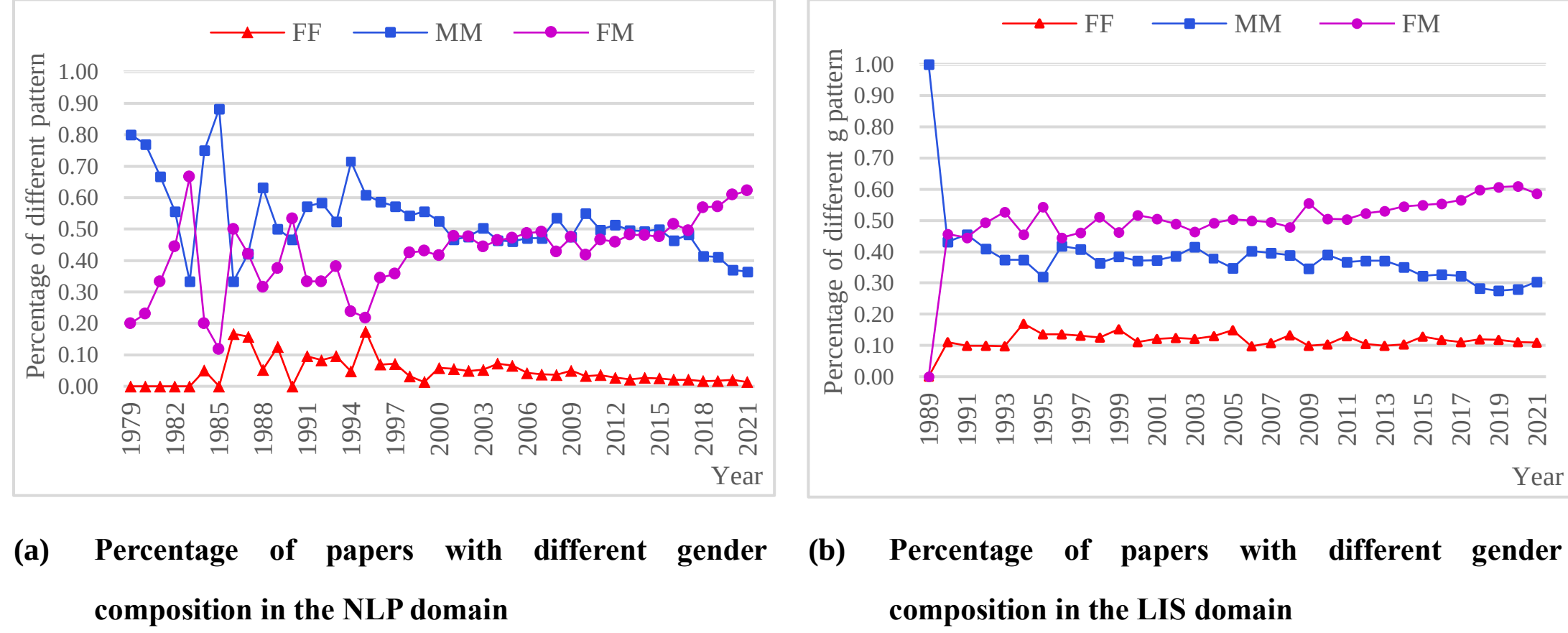

**(a) Percentage of papers with different gender composition in the NLP domain** **(b) Percentage of papers with different gender composition in the LIS domain**

**Figure 4. The percentage of papers with different gender composition by year**

We further investigated the variations in the average B values of papers with different gender collaboration patterns in the domains of NLP and LIS. As shown in Figure 5, the B values for papers under the FF and MM collaboration patterns are both 0, while the average B value for papers under the FM patterns has been decreasing in recent years. This indicates that the gender diversity in collaborative papers within the FM mixed-gender pattern is gradually diminishing.

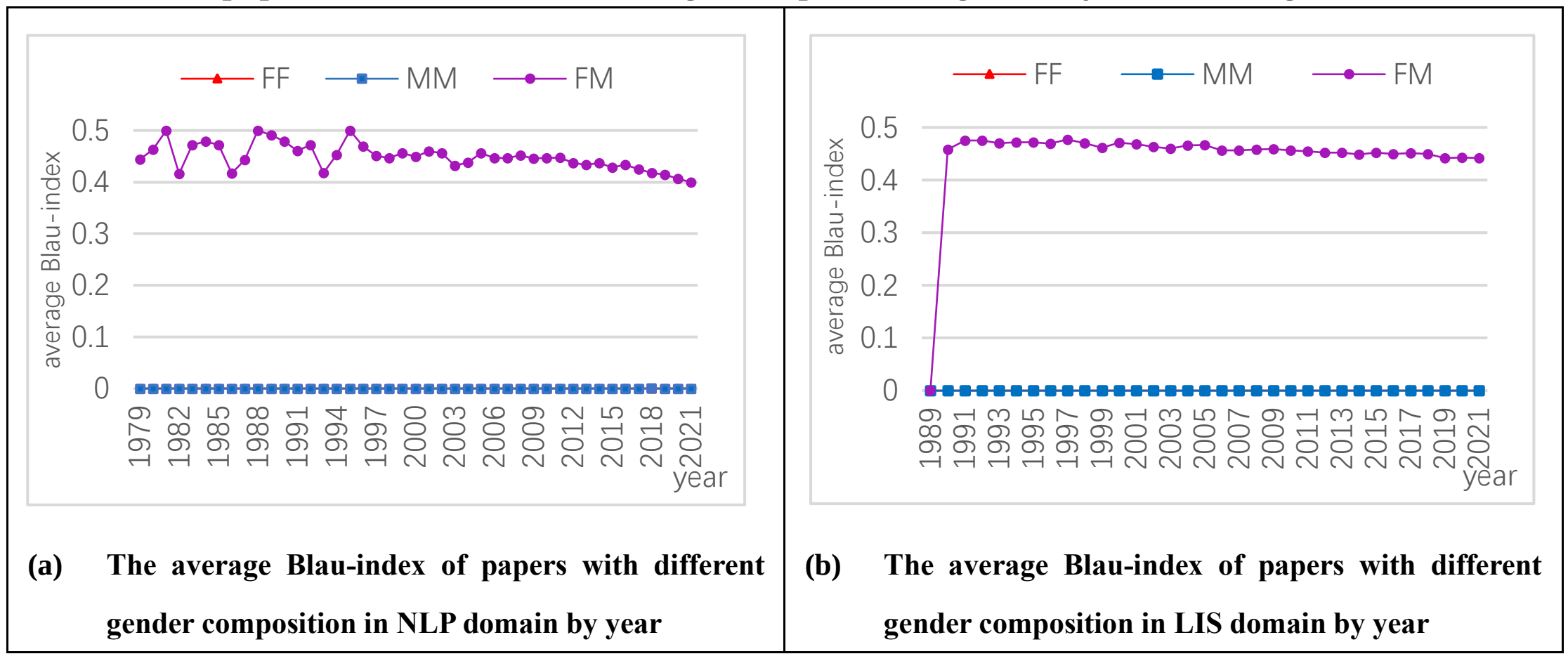

**(a) The average Blau-index of papers with different gender composition in NLP domain by year** **(b) The average Blau-index of papers with different gender composition in LIS domain by year**

**Figure 5. The average Blau-index of papers with different gender composition by year**

### 4.3 Citation analysis of gender composition patterns and gender diversity

To further explore the implications of gender diversity within research teams, this section delves into the relationship between gender diversity in collaboration and the scientific impact of papers.

Additionally, we offer recommendations for achieving a more scientifically balanced gender ratio within team collaborations.

(1) Gender composition patterns and scientific impact of papers

To assess the influence of gender composition patterns on the success of papers, citation counts are commonly employed as a metric. In order to address research question 1 (*RQ1*), we analyze the two-year citation counts of papers with different gender composition year by year.

Figures 6a and 6b respectively display the two-year average citation counts for papers with different gender compositions in the NLP and LIS databases. Figures 7a and 7b respectively display the five-year average citation counts for papers with different gender compositions in the NLP and LIS databases. MM represents all-male collaboration papers, FF represents all-female collaboration papers, and FM represents mixed-gender collaboration papers. It should be noted that the average citation counts of FF collaborative papers display some instability due to their limited quantity. Overall, in our NLP and LIS databases, the average citation counts for MM collaborative papers and FM collaborative papers are higher than those for FF collaborative papers. Figures 6 and 7 show that the number of citations for FM collaborative papers is on the rise. These results indicate differences in the average citation counts of papers composed of different genders. In recent years, the scientific impact of papers with mixed-gender collaboration has been increasing.

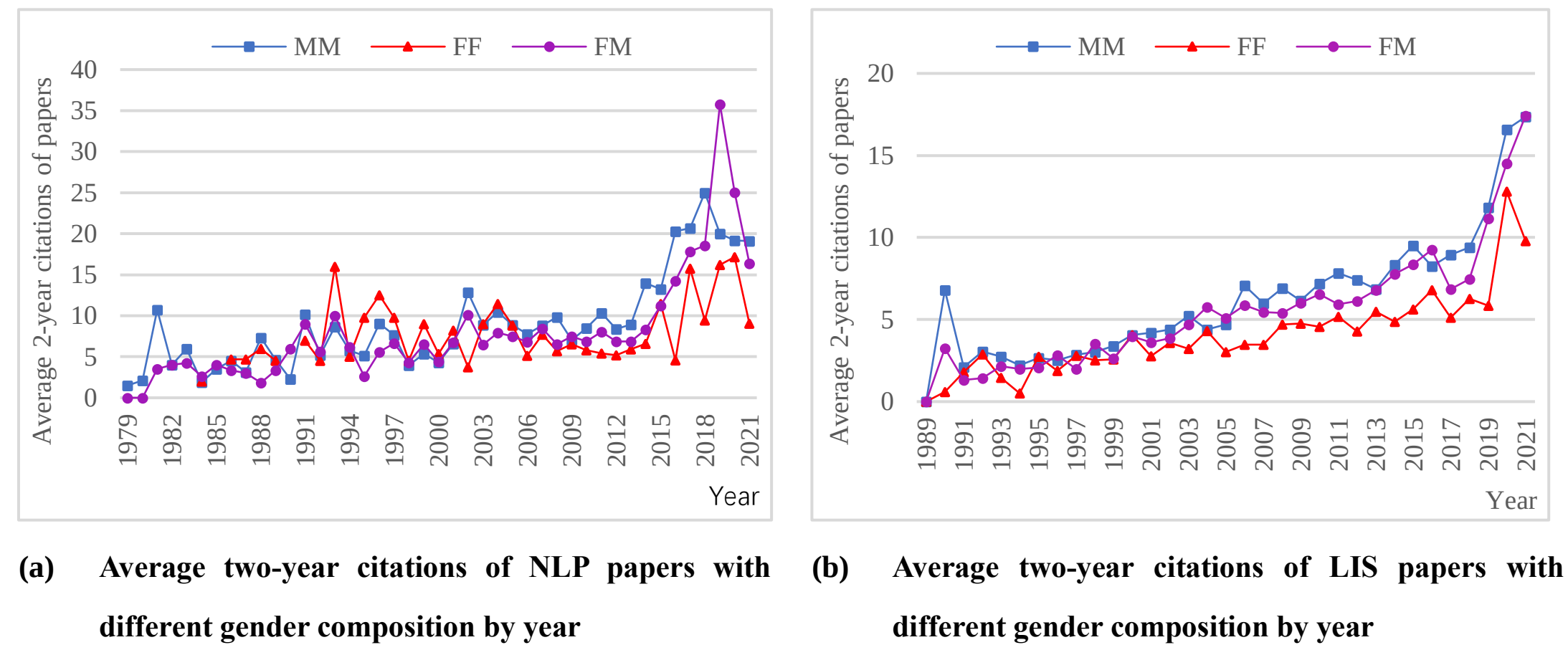

**(a) Average two-year citations of NLP papers with different gender composition by year**

**(b) Average two-year citations of LIS papers with different gender composition by year**

**Figure 6. Average two-year citations of papers with different gender composition by year**

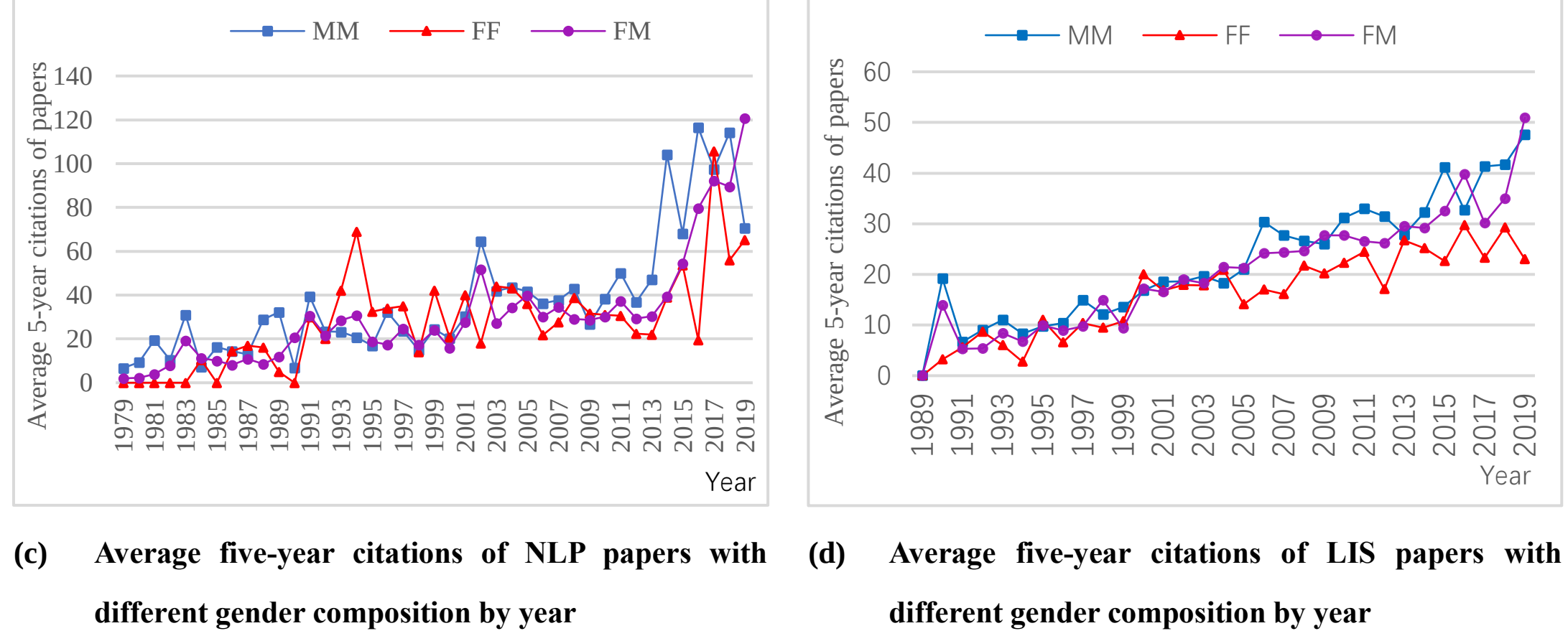

**(c) Average five-year citations of NLP papers with different gender composition by year**

**(d) Average five-year citations of LIS papers with different gender composition by year**

**Figure 7. Average five-year citations of papers with different gender composition by year**

(2) Citation analysis of gender diversity in papers

In recent years, mixed-gender collaboration has become increasingly common in academia, and papers with mixed-gender collaborations have gradually been receiving higher average citation counts compared to those with same-gender collaborations. The extent to which gender diversity characterizes a team could influence the scientific impact of their publications.

To address Research Question 2 (*RQ2*), we scrutinize a total of 14,004 collaborative papers from the top three NLP conferences and 14,113 collaborative papers from the LIS domain (Among 14,528 papers, 415 papers that did not provide institutional information were excluded) to investigate the correlation between gender diversity and the scientific impact of papers. The results of the regression analysis are presented in Table 7.

**Table 7. Regression results of gender diversity and two-year citation counts of papers in the NLP domain**

| **Variables** | **ln (citation+1)** |
|---|---|
| $B^2$ | -2.288*** |
| | (0.482) |
| B | 0.844*** |
| | (0.226) |
| Size | 0.222*** |
| | (0.0282) |
| Ranking | 0.247*** |
| | (0.0224) |
| International | 0.0111 |
| | (0.0240) |
| Institution | -0.0114 |
| | (0.0114) |
| Constant | 0.670 |
| | (0.444) |
| Fixed year effects | YES |
| Fixed topic effects | YES |
| Observations | 14,004 |
| R-squared | 0.130 |

**Note**: *** p<0.01, ** p<0.05, * p<0.1 ( ***, **, * indicate significant at 1%, 5%, and 10% significance levels, respectively)

The fitted equation of the linear regression is shown in formula 3. Based on the fitting function, we conclude that when B = 0.184 (At this point, the proportion of different genders in the team is 89.75% and 10.25%, respectively. In other words, the proportion of male and female authors in the team is 89.75% and 10.25%, respectively, or conversely, the proportion of male and female authors in the team is 10.25% and 89.75%.), papers achieve the highest level of citations. Regression analysis further reveals that team size and institutional ranking have a significant positive correlation with the number of paper citations.

$$\ln\ (citation+1) = -2.288B^2 + 0.844B + 0.222Size + 0.247Ranking \quad (3)$$

Based on the analysis of these 14,113 multi-authored papers in the LIS domain, we examine the relationship between team gender diversity and the number of paper citations. The results of the regression analysis are presented in Table 8.

**Table 8. Regression results of gender diversity and two-year citation counts of papers in the LIS domain**

| VARIABLES | ln (citation+1) |
|---|---|
| $B^2$ | -1.170** |
| | (0.536) |
| B | 0.472* |
| | (0.256) |
| Size | 0.155*** |
| | (0.0415) |
| Ranking | 0.0608*** |
| | (0.0161) |
| International | 0.0905*** |
| | (0.0213) |
| Institution | 0.0294*** |
| | (0.00791) |
| Constant | 0.448 |
| | (0.921) |
| Fixed year effects | YES |
| Fixed topic effects | YES |
| Observations | 14,113 |
| R-squared | 0.146 |

**Note**: *** p<0.01, ** p<0.05, * p<0.1 ( ***, **, * indicate significant at 1%, 5%, and 10% significance levels, respectively)

The fitted equation of the linear regression is shown in formula 4.

$$\ln(\text{citation}+1) = -1.170B^2 + 0.472B + 0.155Size + 0.0608Ranking + 0.0294Institution + 0.0905International \quad (4)$$

In the Library and Information Science domain, it is computed that when B=0.202 (At this point, the proportion of different genders in the team is 88.6% and 11.4%, respectively. In other words, the proportion of male and female authors in the team is 88.6% and 11.4%, respectively, or conversely, the proportion of male and female authors in the team is 11.4% and 88.6%.), the paper achieves the highest citation count. According to the regression results, there still exists an inverted U-shaped relationship between team gender diversity and the number of paper citations. The research findings based on the top three conferences in the NLP domain also hold true in the LIS domain. We find that conference papers in the NLP domain and journal papers in the LIS domain exhibit nearly identical the most impactful team gender diversity.

Correlation analysis indicates an inverted U-shaped relationship between team gender diversity and the number of paper citations. This finding suggests that the positive impact of gender diversity predicted by the information/decision-making perspective interacts with the negative impact predicted by social identity theory and self-categorization theory. An most impactful level of gender diversity exists in paper collaboration. From an information/decision-making perspective, a certain level of gender diversity within a team helps break fixed mindsets, bringing richer knowledge backgrounds, thinking pathways, and problem-solving approaches. In the fields of NLP or LIS, when one gender reaches a ratio of 5-15% within the team, this proportion allows the contributions of minority members to play a role, thereby increasing the breadth and diversity of information in the team's decision-making process. This diversity helps reduce the occurrence of groupthink, making the team more open in the processes of research topic selection, method design, and data analysis interpretation, potentially enhancing the impact of research outcomes.

Social identity theory and self-categorization theory suggest that people often classify members into in-groups and out-groups based on significant social identity characteristics such as gender, race, and nationality. In the fields of NLP or LIS, when the ratio of a particular gender in the team is too low (e.g., below 5%), the perspectives of the minority group are more likely to be overlooked or marginalized, thus failing to effectively promote diversity in research decision-making. Conversely, if the ratio of this gender is too high (e.g., above 15%, yet still a minority), it may form a clearly defined "minority-majority" opposition, leading to more intense conflicts within the team due to differences in identity, reducing collaboration efficiency and research quality. When the minority group constitutes 5-15% of the team, this identity characteristic group has sufficient numbers to avoid being completely marginalized while not forming a distinct subgroup in opposition to the majority. This balance maintains the expansion of thought brought by diversity in a relatively harmonious team atmosphere, without triggering overly strong social identity opposition, reducing conflict costs within the team, and positively contributing to the enhancement of the impact of academic outcomes.

In summary, a gender ratio of 5-15% represents a critical range for the most impactful gender diversity. It provides moderate viewpoint diversity from an information/decision-making perspective while avoiding significant identity differentiation and conflict at the levels of social identity and self-categorization. This balance may be key to enhancing the impact of academic research outcomes. Kanter categorized collaboration into four groups based on gender ratios: uniform group (0/100), skewed group (5/95-15/85), skewed group (20/80-35/65), and balanced group (40/60-50/50) (Kanter, 1977; Kanter, 1987). Our regression model indicates that for NLP papers, a team composition with a proportion of 89.75% and 10.25% for different genders is associated with maximum paper impact. In contrast, for LIS papers, the corresponding proportions of different genders in the team being 88.6% and 11.4% are related to maximum paper impact. According to Kanter's categorization, our study reveals that the best gender diversity for the scientific impact of papers falls within the skewed group with a gender ratio ranging from 5% to 15%. This implies that the greatest collaborative benefit is obtained when the gender ratio within a team lies between 5% and 15%.

### 4.4 Robustness test

In the previous analysis, we identified an inverted U-shaped relationship between the gender diversity of the author teams and the two-year citation counts of the papers. To examine whether this inverted U-shaped relationship persists over a longer citation period, we investigated the relationship between the gender diversity of author teams and the five-year citation counts of papers in the domains of NLP and LIS. Due to the unavailability of five-year citation counts for papers published after 2019, our verification dataset excluded papers published after that year. The regression results for papers in the NLP domain are presented in Table S3, while the regression results for papers in the LIS domain are shown in Table S4. The regression results indicate that the gender diversity of author teams is still associated with an inverted U-shaped relationship regarding the five-year citation counts, with the most impactful range of gender diversity remaining at a proportion of a certain gender within the team being between 5% and 15%.

After removing the extreme years with limited paper data from the dataset, we re-examined the relationship between the gender diversity of author teams and citation counts. The results indicate that for both NLP (Table S5) and LIS data (Table S6), the gender diversity of author teams continues to demonstrate an inverted U-shaped relationship with citation counts even after the removal of extreme data. Furthermore, the most impactful range remains consistent, with the proportion of a certain gender within the team being between 5% and 15%.

This study employs the B-index to measure gender diversity within teams. In this section, we utilize Shannon entropy to assess team gender diversity, with the regression results presented in Table S7 and Table S8. When measuring gender diversity using entropy, the most impactful entropy value for NLP papers is 0.397, which corresponds to a gender ratio of approximately 0.931/0.069 for the team. For LIS papers, the most impactful entropy value is 0.426, corresponding to a gender ratio of approximately 0.911/0.089. The most impactful range for gender proportion remains consistent, with the representation of a certain gender within the team being between 5% and 15%.

## 5. Discussion

We investigate gender differences among authors of academic papers in the natural language processing and Library and Information Science domain. Our results indicate that there is underrepresentation of women in both the NLP and LIS domains, and the gender disparity in the NLP domain is larger compared to the LIS domain. Additionally, we find an inverted U-shaped relationship between the gender diversity of paper collaboration and paper citation counts. As gender diversity increases from 0 (representing gender homogeneity with a 0/100 gender ratio) to 0.5 (representing maximum gender diversity with a 50/50 gender ratio), paper citation counts initially rises and then falls. Our research suggests that the most impactful gender range for teams in NLP and LIS domains is 5-15% for one gender of the total authors.

### 5.1 Comparison with Previous Research

Previous studies have suggested that mixed-gender team collaboration is associated with greater scientific impact of papers (Campbell et al., 2013; Maddi & Gingras, 2021; Yang et al., 2022). However, while the knowledge of this relationship provides valuable insights, it does not offer specific guidance for team formation. The most pertinent finding related to our research is that Yang et al. (2022) discovered that the highest level of gender diversity, characterized by equal representation of males and females, was associated with a greater scientific impact of papers in the medical field. However, our research findings reveal a unique inverted U-shaped relationship between gender diversity within paper teams and the scientific impact of papers in the domains of NLP and LIS. Specifically, we observe that the proportion of one gender within a team ranging from 5/100 to 15/100 is associated with a greater scientific impact of papers. Regarding the reasons for this difference, Larivière et al. (2016) demonstrated differences in the division of labor among team members across different fields, while Larivière et al. (2021) highlighted the existence of gender-related preferences in team roles. These findings indicate that the most impactful range of gender ratios within teams may be influenced by factors such as preferences related to gender roles across different domains. Therefore, it is necessary to investigate the most impactful gender ratio ranges in different fields.

### 5.2 Implications of this study

The study is the first to clearly indicate that in the domains of Natural Language Processing (NLP) and Library and Information Science (LIS), when the proportion of a specific gender within a team falls between 5% and 15%, the team's gender diversity is associated with a higher academic impact of the papers produced by the team. This inverted U-shaped relationship provides a quantitative reference for team building.

In team collaboration, from the perspectives of information/decision-making, social identity theory and self-categorization theory, it is essential to value the diversity brought by authors of different genders in terms of knowledge, skills, and ways of thinking. This diversity not only forms the foundation of team innovation but also provides multi-faceted approaches to solving complex problems. Therefore, in practice, it is necessary to design a well-structured team to foster interaction and communication among members of different genders. At the same time, by establishing an inclusive team culture, misunderstandings or conflicts arising from gender differences can be minimized. Managers should pay attention to the needs and tendencies of members from different genders and adopt appropriate incentive mechanisms to enhance trust and a sense of belonging among members, thereby improving team cohesion and overall performance. Moreover, through scientific team management, the potential of gender diversity should be fully leveraged, providing stronger support for knowledge creation and problem-solving.

Promoting multidisciplinary research areas such as human-centered themes to attract more female researchers. In this study, we find that men dominate both the number of authors and the citation numbers of papers. To promote gender equality in academia, it is important to explore interdisciplinary research areas and adopt qualitative methodologies inclusive of person-centered themes. Moreover, addressing gender disparities in academia requires promoting a broader and more inclusive culture that supports women, eliminating potential career discrimination, emphasizing STEM education for females, and exposing them to traditionally male-dominated fields from an early age.

### 5.3 Limitations of this study

The present study has several limitations that should be recognized.

First, the dataset only includes papers from the NLP and LIS domains. To investigate the relationship between team gender diversity and the scientific impact of papers, we explore the most impactful gender diversity within teams using papers from the top three NLP conferences in the NLP domain and papers from 14 journals in the LIS domain. We find that the most impactful gender diversity for teams in both the LIS and NLP domains is when the proportion of any single gender within the team ranges from 5/100 to 15/100. However, it should be noted that the findings of this study may not fully reflect the gender differences among authors in the broader scientific community. Additionally, the most impactful gender diversity for collaboration in papers may vary across research domains beyond NLP and LIS.

In addition, the regression analysis was conducted without controlling for other factors that may affect scientific impact. In this study, the correlation between team gender diversity and scientific impact of papers is analyzed while controlling for factors such as team size, number of institutions, institutional ranking, international collaboration, publication year, and topic of the paper, which could potentially impact the scientific impact of papers. However, it is important to note that other factors that may influence the scientific impact of papers, such as the average academic age of the team, the nationalities of team members, collaborating institutions, the quantity and quality of papers published by the authors, the authors' collaboration networks, and the patterns of repeated collaborations among authors, have not been taken into account. This limitation may introduce some bias into the research findings.

Finally, we measured scientific impact using only the citation counts of the papers. In our analysis of the scientific impact of papers, we have solely relied on the citation numbers as a measure of paper quality and representation of scientific impact. However, it is crucial to acknowledge that the number of citations may not encompass the entirety of a paper's quality. Other significant indicators, such as the number of times papers are read and the size of the influence of papers' citations, also contribute to assessing the research's excellence. Therefore, relying solely on the number of citations to measure the scientific impact of papers may be a one-sided approach.

## 6. Conclusion and future work

This study investigates the relationship between gender diversity in team collaboration and the scientific impact of papers, using a dataset comprising NLP conference papers from *ACL*, *NAACL*, and *EMNLP*, as well as papers from 14 journals in the LIS domain. The main findings are as follows:

Firstly, we identify differences in the gender composition patterns of papers. Collaborative papers in both the NLP and LIS domains are increasingly common, with mixed-gender collaborations comprising the highest percentage. These mixed-gender collaborative papers gradually receive higher average citation counts compared to same-gender collaborative papers. Secondly, we determine the most impactful interval for gender ratio in team collaboration. The relationship between gender diversity in paper collaboration and the number of citations follows an inverted U-shape. We find that the most impactful gender balance in NLP and LIS teams is 5% to 15% for one gender among the total authors. Finally, our research indicates that in the domains of NLP and LIS, women are underrepresented among paper authors, first authors, and last authors. Furthermore, the gender gap is more significant in the NLP domain compared to the LIS domain.

There are several areas for potential improvement in future studies. Firstly, future research could be expanded to a broader range of domains. It could investigate whether gender diversity in papers across different domains is related to the impact of the papers, and whether there exists a corresponding most impactful gender diversity. Secondly, it may be beneficial to control for other factors in the analysis of the correlation between gender ratios and the scientific impact of papers, such as the average academic age of the team, the publication records of the authors, and the authors' collaborative networks. Additionally, while we identify a range of team gender ratios that are associated with higher scientific impact of papers, this study does not reveal the underlying causal relationship. Future research can further explore the reasons behind the differences in the scientific impact of papers among teams with varying author gender ratios. Finally, measuring the scientific impact of paper solely based on the number of citations may not comprehensively reflect the quality of a paper. Future studies could consider incorporating other Altmetrics indices, such as the number of times a paper is downloaded or read, or the influence of the paper's citations, to more comprehensively measure the impact of a paper.

## Acknowledgements

This work was supported by the National Natural Science Foundation of China (Grant No.72074113) and Key Project of Jiangsu Provincial Social Science Foundation (Grant No. 20TQA001).

## Appendix

**Table S1. Correlation coefficient analysis of variables in NLP papers**

| | ln(citation+1) | B | Year | Size | Institution | International | Topic | Ranking |
|---|---|---|---|---|---|---|---|---|
| ln(citation+1) | 1.000 | | | | | | | |
| B | -0.007 | 1.000 | | | | | | |
| Year | 0.262*** | 0.090*** | 1.000 | | | | | |
| Size | 0.145*** | 0.101*** | 0.206*** | 1.000 | | | | |
| Institution | 0.094*** | 0.093*** | 0.207*** | 0.341*** | 1.000 | | | |
| International | 0.054*** | 0.043*** | 0.112*** | 0.131*** | 0.530*** | 1.000 | | |
| Topic | -0.166*** | -0.122*** | -0.403*** | -0.146*** | -0.127*** | -0.044*** | 1.000 | |
| Ranking | 0.124*** | 0.038*** | 0.098*** | 0.083*** | 0.180*** | 0.151*** | -0.061 | 1.000 |

**Note**: *** p<0.01, ** p<0.05, * p<0.1 ( ***, **, * indicate significant at 1%, 5%, and 10% significance levels, respectively)

**Table S2. Correlation coefficient analysis of variables in LIS papers**

| | ln(citation+1) | B | Year | Size | Institution | International | Topic | Ranking |
|---|---|---|---|---|---|---|---|---|
| ln(citation+1) | 1.000 | | | | | | | |
| B | 0.003 | 1.000 | | | | | | |
| Year | 0.310*** | 0.061*** | 1.000 | | | | | |
| Size | 0.068*** | 0.085*** | 0.083*** | 1.000 | | | | |
| Institution | 0.132*** | 0.062*** | 0.185*** | 0.235*** | 1.000 | | | |
| International | 0.124*** | 0.011 | 0.156*** | 0.081*** | 0.495*** | 1.000 | | |
| Topic | -0.016* | -0.005 | 0.060*** | 0.006 | 0.005 | 0.006 | 1.000 | |
| Ranking | 0.058*** | 0.010 | 0.010 | 0.046*** | 0.204*** | 0.230*** | 0.038*** | 1.000 |

**Note**: *** p<0.01, ** p<0.05, * p<0.1 ( ***, **, * indicate significant at 1%, 5%, and 10% significance levels, respectively)

**Table S3. Regression results of gender diversity and five-year citation counts of papers in the NLP domain**

| VARIABLES | ln (citation+1) |
|---|---|
| B2 | -3.483*** |
| | (0.695) |
| B | 1.333*** |
| | (0.327) |
| size | 0.296*** |
| | (0.0441) |
| ranking | 0.309*** |
| | (0.0293) |
| international | 0.0182 |
| | (0.0334) |
| institution | -0.0281* |
| | (0.0166) |
| Constant | 1.662*** |
| | (0.518) |
| Fixed year effects | YES |
| Fixed topic effects | YES |
| Observations | 10,486 |
| R-squared | 0.132 |

**Note**: *** p<0.01, ** p<0.05, * p<0.1 ( ***, **, * indicate significant at 1%, 5%, and 10% significance levels, respectively)

**Table S4. Regression results of gender diversity and five-year citation counts of papers in the LIS domain**

| VARIABLES | ln (citation+1) |
|---|---|
| B2 | -1.195* |
| | (0.631) |
| B | 0.482 |
| | (0.301) |
| size | 0.0978* |
| | (0.0503) |
| ranking | 0.109*** |
| | (0.0184) |
| international | 0.103*** |
| | (0.0248) |
| institution | 0.0151 |
| | (0.00969) |
| Constant | 0.739 |
| | (0.981) |
| Fixed year effects | YES |
| Fixed topic effects | YES |
| Observations | 12,188 |
| R-squared | 0.179 |

**Note**: *** p<0.01, ** p<0.05, * p<0.1 ( ***, **, * indicate significant at 1%, 5%, and 10% significance levels, respectively)

**Table S5. Regression results of gender diversity and two-year citation counts of papers in the NLP domain from 2000 to 2021**

| VARIABLES | ln (citation+1) |
|---|---|
| B2 | -2.259*** |
| | (0.491) |
| B | 0.826*** |
| | (0.230) |
| size | 0.221*** |
| | (0.0285) |
| ranking | 0.247*** |
| | (0.0233) |
| international | 0.0100 |
| | (0.0244) |
| institution | -0.00957 |
| | (0.0116) |
| Constant | 1.619*** |
| | (0.0931) |
| Fixed year effects | YES |
| Fixed topic effects | YES |
| Observations | 13,366 |
| R-squared | 0.112 |

**Note**: *** p<0.01, ** p<0.05, * p<0.1 ( ***, **, * indicate significant at 1%, 5%, and 10% significance levels, respectively)

**Table S6. Regression results of gender diversity and two-year citation counts of papers in the LIS domain from 2000 to 2021**

| VARIABLES | ln (citation+1) |
|---|---|
| B2 | -1.194** |
| | (0.561) |
| B | 0.482* |
| | (0.267) |
| size | 0.155*** |
| | (0.0427) |
| ranking | 0.0372** |
| | (0.0172) |
| international | 0.0963*** |
| | (0.0222) |
| institution | 0.0315*** |
| | (0.00818) |
| Constant | 1.530*** |
| | (0.0662) |
| Fixed year effects | YES |
| Fixed topic effects | YES |
| Observations | 12,704 |
| R-squared | 0.109 |

**Note**: *** p<0.01, ** p<0.05, * p<0.1 ( ***, **, * indicate significant at 1%, 5%, and 10% significance levels, respectively)

**Table S7. Regression results of gender diversity and two-year citation counts of papers in the NLP domain**

**(Entropy measures gender diversity)**

| VARIABLES | ln (citation+1) |
|---|---|
| Entropy 2 | -0.704*** |
| | (0.141) |
| Entropy | 0.559*** |
| | (0.133) |
| size | 0.218*** |
| | (0.0284) |
| ranking | 0.247*** |
| | (0.0224) |
| international | 0.0111 |
| | (0.0240) |
| institution | -0.0111 |
| | (0.0114) |
| Constant | 0.670 |
| | (0.444) |
| Fixed year effects | YES |
| Fixed topic effects | YES |
| Observations | 14,004 |
| R-squared | 0.130 |

**Note**: *** p<0.01, ** p<0.05, * p<0.1 ( ***, **, * indicate significant at 1%, 5%, and 10% significance levels, respectively)

**Table S8. Regression results of gender diversity and two-year citation counts of papers in the LIS domain (Entropy measures gender diversity)**

| VARIABLES | ln (citation+1) |
|---|---|
| Entropy 2 | -0.370** |
| | (0.163) |
| Entropy | 0.315** |
| | (0.157) |
| size | 0.153*** |
| | (0.0416) |
| ranking | 0.0608*** |
| | (0.0161) |
| inter | 0.0905*** |
| | (0.0213) |
| insistution | 0.0295*** |
| | (0.00791) |
| Constant | 0.448 |
| | (0.921) |
| Fixed year effects | YES |
| Fixed topic effects | YES |
| Observations | 14,113 |
| R-squared | 0.146 |

**Note**: *** p<0.01, ** p<0.05, * p<0.1 ( ***, **, * indicate significant at 1%, 5%, and 10% significance levels, respectively)

# Supplemental Material

The supplemental material primarily details the author gender recognition methods employed, the precision of each gender recognition method, and the number of authors identified by each method.

## 1. Methods for Author Gender Identification

In this paper, we use Mohammad (2020a) approach, which utilizes name gender lists for author gender identification. This approach includes the use of the author gender list provided in the Vogel and Jurafsky (2012) (VJ-AA list), the first name gender list generated by the U.S. Social Security Administration (USSA list), the name gender list obtained from the PUBMED authors and their gender lists (PUBMED list) (Mohammad, 2020a), as well as the World Gender Name Dictionary (WGND) developed by Raffo and Lax-Martinez. For authors whose names cannot be identified through these gender lists, we utilize the Genderize.io tool for gender identification. Finally, if the gender of authors cannot be determined through the previous methods, we resort to using search engine for identification. The process of gender identification is illustrated in Figure 1.

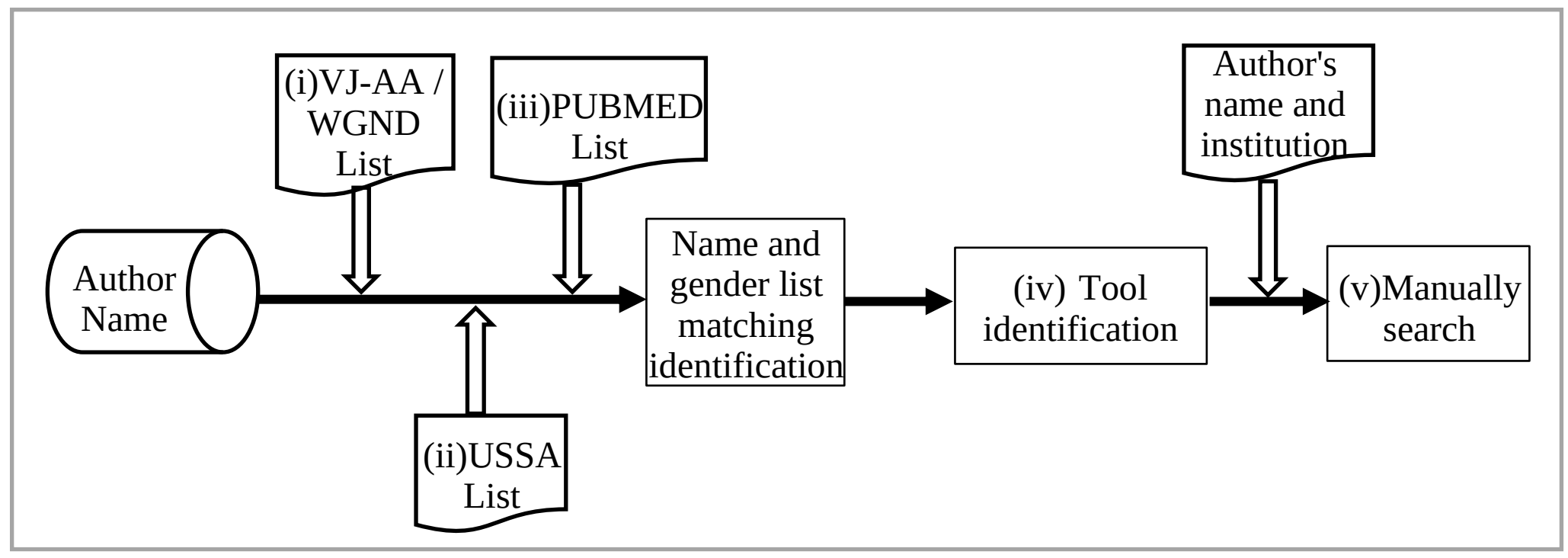

**Figure 1. Flowchart of author gender identification**

The gender identification methods and resources used in this paper are as follows:

(i) Matching with VJ-AA List or WGND List: The VJ-AA list is a compilation of 11,932 authors and their genders in the ACL anthology provided by Vogel and Jurafsky (2012) (referred to as VJ-AA), which includes information on 3,359 women and 8,573 men[3], therefore, when identifying the gender of NLP papers, we check if the author's full name matches any entries in the VJ-AA list. For gender identification in LIS papers, we initially utilize the World Gender Name Dictionary (WGND) developed by Raffo and Lax-Martinez and made available on the Harvard Dataverse[4].

(ii) Matching with USSA List: We verify if the author's first name aligns with any entries in the USSA list. The USSA list consists of 60,182 first names predominantly associated with females and 33,883 first names predominantly associated with males. This list is generated from the U.S. Social Security Administration's (USSA) published database of newborn first names and genders[5].

[3] https://nlp.stanford.edu/projects/gender.shtml
[4] https://doi.org/10.7910/DVN/YPRQH8
[5] https://www.ssa.gov/oact/babynames/limits.html

(iii) Matching with PUBMED List: We determine whether the author's first name corresponds to any entries in the PUBMED list. The PUBMED list encompasses 22,693 first names closely associated with females and 29,531 first names closely associated with males. This data is obtained from a list of 9,300,182 PUBMED[6] authors and their genders[7] (Torvik & Smalheiser, 2009; Smith et al., 2013).

(iv) Utilizing Genderize.io[8] and Genni 2.0 Tool: We employ the Genderize.io and Genni 2.0 tool to assign gender to the author. Genderize.io is an automated gender identification tool that utilizes a database of names constructed from publicly available census statistics. This tool is provided by Demografix ApS. Genni 2.0 Tool is provided by Smith et al. (2013).

(v) Internet Search: If the author's gender cannot be identified through the previous methods, we conduct an internet search using search engines such as Google (Google.com), Microsoft Bing (Bing.com), and others. Through this search, we check the author's affiliation and personal page to identify the author's gender.

The gender associations of first names are determined using the PUBMED and USSA name-gender lists. This calculation involves assessing the percentages of first names that correspond to male and female genders. If a first name has a percentage higher than 95%, it is considered strongly associated with a particular gender (Mohammad, 2020a). In addition, the Genderize.io and Genni 2.0 tool relies on a database of names constructed from public census statistics and other statistical data to determine gender. When a first name is entered into the tool, it provides information on the probability of identifying the name as a specific gender.[9] For this study, we establish criteria for Genderize.io and Genni 2.0 tool based on its precision and recall in identifying the gender of authors in the VJ-AA list. Specifically, we consider a name's gender to be determined by Genderize.io if it has a probability of appearing in the database of five or more times and is associated with a certain gender at a probability of 0.7 or higher. When the Genni 2.0 Tool displays a prediction accuracy higher than 85%, we consider the identified gender as the author's gender.

## 2. Gender identification precision and identification counts by different methods

In order to assess the accuracy of gender identification using the WGND List, USSA list, PUBMED list, and Genderize.io, we employ these methods to predict the gender of authors in the VJ-AA list. The accuracy of author gender prediction using various methods is evaluated by comparing their prediction results to the manually determined gender in the VJ-AA list. We evaluate the performance of gender identification using Precision (P) indicators, and provide the precision of each method in recognizing different genders. P-M, P-F, and P-ALL respectively represent the precision rates of identifying male authors, female authors, and all authors.

$$P = \frac{Number\ of\ correct\ gender\ identified}{Total\ number\ of\ genders\ identified} \tag{1}$$

[6] https://pubmed.ncbi.nlm.nih.gov/
[7] https://experts.illinois.edu/en/datasets/genni-ethnea-for-the-author-ity-2009-dataset-2
[8] https://genderize.io/
[9] https://genderize. io/#overview

Table 1 presents the precision (P) values of author gender identification using various methods. The results indicate that the gender recognition precision of various methods overall exceeds 90%, demonstrating the reliability of the gender recognition results in this study.

**Table 1. Gender identification precision of different methods**

| Indicators / Resource | P-M (%) | P-F (%) | P-ALL (%) |
|---|---|---|---|
| **USSA List** | 98.92 | 97.08 | 98.42 |
| **PUBMED List** | 98.76 | 94.96 | 97.75 |
| **WGND List** | 99.22 | 96.40 | 98.35 |
| **Genderize.io** | 98.13 | 88.83 | 94.85 |
| **Genni 2.0** | 99.16 | 96.44 | 98.45 |

NLP dataset consists of 15,337 conference papers, involving a total of 52,541 authors and encompassing 17,122 unique first names[10]. LIS dataset consists of 23,957 papers, involving a total of 57,052 authors and encompassing 35,166 unique first names. To ensure accurate gender identification, we employ a combination of name-gender list matching identification, the Genderize.io and Genni 2.0 tool, and manual search identification (Zhang et al., 2023a). The gender of all authors in NLP papers has been determined using these methods. For LIS papers, due to the presence of abbreviations in the names of many authors, making it impossible to identify their gender, we excluded 2,057 papers with incomplete gender recognition. The remaining dataset consists of 21,900 papers. In the end, we identified the gender of 17,122 names in the NLP domain and 32,335 names in the LIS domain. The number of authors' genders identified by different methods is shown in Table 2.

**Table 2. Number of authors identified by different methods.**

| Domain / Method | NLP | LIS |
|---|---|---|
| **USSA / PUBMED / WGND/ VJ-AA List** | 13,126 (76.66%) | 26,319 (81.39%) |
| **Genderize.io/ Genni 2.0** | 1,792 (10.47%) | 3,786 (11.71%) |
| **Internet Search** | 2,204 (12.87%) | 2,230 (6.90%) |
| **Total** | 17,122 (100%) | 32,335 (100%) |